\documentclass[aps,twocolumn,superscriptaddress,floatfix,nofootinbib,showpacs,amsmath,amssymb,altaffilletter]{revtex4-1}
\pdfoutput=1
\usepackage{graphicx}
\usepackage{bm}
\usepackage{subfigure}
\usepackage{url}
\usepackage[hyperindex]{hyperref}
\usepackage{color}
\usepackage[ddmmyy,24hr]{datetime}
\usepackage{bigdelim}
\usepackage{booktabs}
\usepackage{dcolumn}
\usepackage{multirow}
\usepackage{subfigure}
\usepackage{cancel}
\usepackage{stackrel}
\usepackage{paralist}
\usepackage{xspace}
\usepackage[utf8]{inputenc}
\newcommand{\nua}[1]{\ensuremath{\rlap{\kern-2.5pt\ensuremath{\overset{\scriptscriptstyle(-)}{\phantom{\nu}}}}{\ensuremath{{\nu}_{#1}}}}}

\newcommand{\beq}{\begin{equation}}
\newcommand{\eeq}{\end{equation}}
\newcommand{\bea}{\begin{eqnarray}}
\newcommand{\eea}{\end{eqnarray}}

\newcommand{\eps}{\varepsilon}

\newcommand{\mzd}{m_{Z_d}}

\usepackage{enumitem}

\makeatletter
\def\namedlabel#1#2{\begingroup
    #2%
    \def\@currentlabel{#2}%
    \phantomsection\label{#1}\endgroup
}

\makeatother
\begin{document}

\title{Muon and electron g-2 and proton and cesium weak charges implications on dark $\mathbf{Z_d}$ models}

\author{M.~Cadeddu}
\email{matteo.cadeddu@ca.infn.it}
\affiliation{Istituto Nazionale di Fisica Nucleare (INFN), Sezione di Cagliari,
Complesso Universitario di Monserrato - S.P. per Sestu Km 0.700,
09042 Monserrato (Cagliari), Italy}

\author{N.~Cargioli}
\email{nicola.cargioli@ca.infn.it}
\affiliation{Dipartimento di Fisica, Universit\`{a} degli Studi di Cagliari,
Complesso Universitario di Monserrato - S.P. per Sestu Km 0.700,
09042 Monserrato (Cagliari), Italy}
\affiliation{Istituto Nazionale di Fisica Nucleare (INFN), Sezione di Cagliari,
Complesso Universitario di Monserrato - S.P. per Sestu Km 0.700,
09042 Monserrato (Cagliari), Italy}

\author{F.~Dordei}
\email{francesca.dordei@cern.ch}
\affiliation{Istituto Nazionale di Fisica Nucleare (INFN), Sezione di Cagliari,
Complesso Universitario di Monserrato - S.P. per Sestu Km 0.700,
09042 Monserrato (Cagliari), Italy}

\author{C.~Giunti}
\email{carlo.giunti@to.infn.it}
\affiliation{Istituto Nazionale di Fisica Nucleare (INFN), Sezione di Torino, Via P. Giuria 1, I--10125 Torino, Italy}

\author{E.~Picciau}
\email{emmanuele.picciau@ca.infn.it}
\affiliation{Dipartimento di Fisica, Universit\`{a} degli Studi di Cagliari,
Complesso Universitario di Monserrato - S.P. per Sestu Km 0.700,
09042 Monserrato (Cagliari), Italy}
\affiliation{Istituto Nazionale di Fisica Nucleare (INFN), Sezione di Cagliari,
Complesso Universitario di Monserrato - S.P. per Sestu Km 0.700,
09042 Monserrato (Cagliari), Italy}

%\date{\dayofweekname{\day}{\month}{\year} \ddmmyydate\today, \currenttime}

\begin{abstract}
Theories beyond the standard model involving a sub-GeV-scale vector $Z_d$ mediator have been largely studied as a possible explanation of the experimental values of the muon and electron anomalous magnetic moments. Motivated by the recent determination of the anomalous muon magnetic moment performed at Fermilab, we derive the constraints on such a model obtained from the magnetic moment determinations and the measurements of the proton and cesium weak charge, $Q_W$, performed at low-energy transfer. In order to do so, we revisit the determination of the cesium $Q_W$ from the atomic parity violation experiment, which depends critically on the value of the average neutron rms radius of $^{133}\text{Cs}$, by determining the latter from a practically model-independent extrapolation from the recent average neutron rms radius of $^{208}\text{Pb}$ performed by the PREX-2 Collaboration. From a combined fit of all the aforementioned experimental results, we obtain rather precise limits on the mass and the kinetic mixing parameter of the $Z_d$ boson, namely $m_{Z_d} = 47{}^{+61}_{-16} \, \mathrm{MeV}$
and $\varepsilon = 2.3{}^{+1.1}_{-0.4} \times 10^{-3}$, when marginalizing over the $Z-Z_d$ mass mixing parameter $\delta$.

\end{abstract}

\maketitle

A new measurement of the anomalous muon magnetic moment, referred to as $a_\mu \equiv (g_\mu-2)/2$, has been largely awaited due to the presence of a long-standing deviation of the experimental determination of $a_\mu$, performed at BNL~\cite{Bennett_2006} in 2004, from the theoretical expectation of about $3.7\sigma$.
Recently, the Muon g-2 Collaboration at Fermilab (FNAL) released a new measurement~\cite{PhysRevLett.126.141801}, with a slightly better precision, about $15\%$ less, than the BNL one, which is
$a_\mu^{\rm FNAL,\,exp} = 116\ 592\ 040(54) \times 10^{-11}$.
The combined experimental average between the FNAL and BNL results 
\beq
a_\mu^{\rm exp} = 116\ 592\ 061(41) \times 10^{-11},
\label{amuexp}
\eeq
can be compared with the standard model (SM) prediction $a_\mu^{\rm SM} = 116\ 591\ 810 (43) \times 10^{-11}$~\cite{Aoyama:2020ynm,Aoyama:2012wk,Aoyama:2019ryr,Czarnecki:2002nt,Gnendiger:2013pva,Davier:2017zfy,Keshavarzi:2018mgv,Colangelo:2018mtw,Hoferichter:2019gzf,Davier:2019can,Keshavarzi:2019abf,Melnikov:2003xd,Masjuan:2017tvw,Colangelo:2017fiz,Hoferichter:2018kwz,Gerardin:2019vio,Bijnens:2019ghy,Colangelo:2019uex,Pauk:2014rta,Danilkin:2016hnh,Jegerlehner:2017gek,Knecht:2018sci,Eichmann:2019bqf,Roig:2019reh,Blum:2019ugy,Colangelo:2014qya},
showing an intriguing $4.2 \sigma$ discrepancy 
\beq
\Delta a_\mu = a_\mu^{\rm exp} - a_\mu^{\rm SM} = 251(59) \times 10^{-11} \label{eq:deltaamu}.
\eeq
This breakthrough result strengthens the motivation for the development of SM extensions, in particular in light of other increasing evidences for the incompleteness of the SM recently reported~\cite{Aaij:2021vac}. \\

In the last years, also the electron anomalous magnetic moment experimental result~\cite{Hanneke_2008, Hanneke_2011} has shown a greater than $2\sigma$ discrepancy with the SM prediction~\cite{Parker_2018}, even if with an opposite sign with respect to the muon one. 
However, a new determination of the fine structure constant~\cite{Morel:2020dww}, obtained from the measurement of the recoil velocity on rubidium atoms, resulted into a reevaluation of the SM electron magnetic moment, bringing to a positive discrepancy of about $1.6\sigma$. Namely
\beq
\Delta a_e = a_e^{\rm exp} - a_e^{\rm SM,\, Rb} = 0.48 (30) \times 10^{-12} \label{eq:deltaae},
\eeq
where $a_e\equiv(g_e-2)/2$.
Interestingly, now the electron and muon magnetic moment discrepancies point to the same direction.  \\

These  longstanding  anomalies  have  motivated  a  variety  of  theoretical  models that predict the existence of yet to be discovered particles that might contribute to the process~\cite{Fayet_2007,Davoudiasl_2018,Cadeddu:2020nbr, B_hm_2004, Giudice_2012,Bodas:2021fsy}.
In particular, they could indicate the presence of an additional sub-GeV-scale gauge boson, referred to as $Z_d$~\cite{Davoudiasl:2012ag,Davoudiasl:2012ig,Davoudiasl:2013aya,arcadi2019new}.
Here, we recall the basic features of such a model in which we assume a $U(1)_d$ gauge symmetry associated with a hidden dark sector.  The corresponding $Z_d$ gauge boson couples to the SM bosons via kinetic mixing, parametrized by $\eps$, and $Z$-$Z_d$ mass matrix mixing, parametrized by $\eps_Z = (\mzd/m_Z)\delta$~\cite{Davoudiasl:2012ag}, where $\mzd$ and $m_Z$ are the $Z_d$ and $Z$ masses, respectively.
The parameter $\delta$ in the latter relation is usually replaced~\cite{Davoudiasl_2015} by the following expression
\beq
\delta' \simeq \delta + \frac{\mzd}{m_Z} \,\eps \, \tan\theta_W,
\label{eq:delta'}
\eeq
that incorporates higher order corrections, even if small for $\mzd \ll m_Z$. Here, $\theta_W$ is the SM predicted running of the Weinberg angle in the modified minimal subtraction ($\overline{\mathrm{MS}}$) renormalization scheme~\cite{Zyla:2020zbs,Erler:2004in,Erler:2017knj}.

As a consequence of the mixing, the $Z_d$ coupling with the SM results into an interaction Lagrangian~\cite{Davoudiasl:2012ag, Davoudiasl:2012ig,Davoudiasl:2013aya}
\beq
{\cal L}_{\text{int}} = (- e \eps J^{em }_\mu - \frac{g}{2 \cos \theta_W}\frac{\mzd}{m_Z} \delta' J^{NC}_\mu  ) Z_d^\mu,
\label{Lint}
\eeq
where $e$ is the electric charge, $ J_\mu^{NC}$ and  $J_\mu^{em}$ are respectively the neutral and electromagnetic currents, whereas $Z_d^\mu$ is the new boson field.\\
Within this model, the new weak neutral current amplitudes at low $Q^2$ momentum transfer can be retrieved through the substitutions $G_F\to\rho_d G_F$, $G_F$ being the Fermi coupling constant, and $\sin^2\theta_W(Q^2)\to\kappa_d \, \sin^2\theta_W(Q^2)$~\cite{Davoudiasl:2012ag,Davoudiasl_2015,Davoudiasl:2012qa,Davoudiasl:2014kua}, where
\begin{align}
\rho_d &=1+ (\delta + \frac{\mzd}{m_Z} \,\eps \, \tan\theta_W)^2f\Big(\frac{Q^2}{\mzd^2}\Big),&
\label{rhod}
\end{align}
and
\begin{align}
\kappa_d &=1- \eps (\delta + \frac{\mzd}{m_Z} \,\eps \, \tan\theta_W)  \frac{m_Z}{\mzd} \cot\theta_W f\Big(\frac{Q^2}{\mzd^2}\Big).&\label{kappad}
\end{align}
The term $f({Q^2}/{\mzd^2})$ is related to the propagator of the new boson and it may assume different forms depending on the experimental process~\cite{BOUCHIAT198373,Bouchiat_2005}.\\

The one-loop vector contribution to the magnetic moment of the lepton $l=e,\,\mu$ which arises from this model is~\cite{PhysRevLett.109.031802}
\begin{align}
&a_{l,\, \text{vector}}^{Z_d} = \frac{\alpha}{2\pi} \Big(\eps+\frac{\mzd}{m_Z}\delta' \frac{1-4\sin^2\theta_W}{4\sin\theta_W\cos\theta_W}\Big)^2 F_V\Big(\frac{m_{Z_d} } {m_l }\Big) \label{eq:5},&
\end{align}
where $\sin\theta_W$ is employed at the corresponding lepton mass scale, $\alpha$ is the fine-structure constant, $m_l$ the lepton mass and
\begin{align}
 &F_V(x)\equiv \int_0^1 dz \frac{2 z (1-z)^2}{(1-z)^2 + x^2 z}. \label{eq:6}&
\end{align}
The mass mixing introduces also an axial contribution, which is although negligible, given by~\cite{PhysRevLett.109.031802}
\begin{align}
&a_{l,\,\text{axial}}^{Z_d} = -\frac{G_F m_l^2}{8 \sqrt{2} \pi^2} \delta'^2 F_A \Big(\frac{m_{Z_d} } {m_l }\Big),& \label{eq:7}
\end{align}
where
\begin{align}
&F_A(x) \equiv \int_0^1 dz \frac{2 (1-z)^3 + x^2 z (1-z) (z+3)}{(1-z)^2 + x^2 z}.& \label{eq:8}
\end{align}
Adding the two contributions in Eqs.~(\ref{eq:5}) and (\ref{eq:7}), it is possible to retrieve the total $Z_d$ induced magnetic momentum contribution
$
a_{l}^{Z_d}(\eps,\,\delta,\,\mzd)=a_{l,\,\text{vector}}^{Z_d}+a_{l,\,\text{axial}}^{Z_d} .
$
\\

Another consequence of the existence of this additional $Z_d$ boson, besides the modification of the lepton magnetic moment, would be the introduction of a new source of parity violation that could be tested by experiments sensitive to the weak charge, $Q_W$, of both protons and nuclei. In particular, recently the ${\rm Q_{weak}}$ Collaboration at JLAB~\cite{Androic:2018kni} measured the proton weak charge at $Q^2=0.0248 \,\mathrm{GeV^2}$ to be 
\begin{equation}
    Q_W^{p,\,\rm exp}=0.0719(45),
    \label{Qprotonexp}
\end{equation}
which has to be compared with the SM prediction~\cite{Erler:2013xha,Zyla:2020zbs} that, taking into account radiative corrections, is
\begin{equation}
    Q_W^{p,\,\rm SM}=-2 g^{ep}_{AV}(\sin^2\theta_W)\Big(1-\frac{\alpha}{2\pi}\Big)=0.0711(2),
\end{equation}
where $g^{ep}_{AV}$ is the SM electron-proton coupling, which depends on the weak mixing angle at the appropriate experimental energy scale.\\
Similarly, in the low-energy sector, atomic parity violation (APV) experiments provide the measurement of the weak charge of a nucleus $\mathcal{N}$ with $N$ neutrons and $Z$ protons, which is also very sensitive to new vector bosons. So far, the most precise measurement has been performed at $Q\approx 2.4\ \mathrm{MeV}$ using cesium atoms ($N_{\mathrm{Cs}}=78$ and $Z_{\mathrm{Cs}}=55$), for which one can derive the following SM prediction\footnote{The SM prediction for the weak charge of a nucleus at tree-level is given by $Q_W^{\mathrm{\mathcal{N},\,\ tree}}=-N+ Z(1-4 \sin^2\theta_W)$.}~\cite{Zyla:2020zbs} which also includes radiative corrections
\begin{align}
    Q_W^{\mathrm{^{133}Cs},\, \mathrm{SM}}&=-2\, [\,Z_{\mathrm{Cs}}(g^{ep}_{AV}(\sin^2\theta_W)+0.00005)&\nonumber\\\nonumber
    &+N_{\mathrm{Cs}}(g^{en}_{AV}+0.00006)]\Big(1-\frac{\alpha}{2\pi}\Big)&\\
    &=-73.23(1),&
\end{align}
where $g^{en}_{AV}$ is the SM electron-neutron coupling\footnote{The SM prediction for the electron-proton and electron neutron-couplings at tree-level are given by $g^{ep}_{AV}= -1/2+2\sin^2\theta_W$ and $g^{en}_{AV}= 1/2$.}.
The current experimental measurement~\cite{Wood:1997zq, Guena:2004sq,Zyla:2020zbs}
\begin{align}
    Q_W^{\mathrm{^{133}Cs},\, \mathrm{PDG}}=-72.82(42),&
    \label{APVPDG}
\end{align}
depends strongly on the value of the average neutron rms radius of $^{133}\text{Cs}$, $R_n(^{133}\text{Cs})$~\cite{Pollock1999, PhysRevC.46.2587, Horowitz2001}. Since at the time of Refs.~\cite{Dzuba:2012kx,PhysRevA.65.012106} there was not any cesium neutron radius measurement, the correction on $Q_W^{\mathrm{^{133}Cs},\ \mathrm{PDG}}$ due to the difference between $R_n(^{133}\text{Cs})$ and $R_p(^{133}\text{Cs})$, the average proton rms radius, could only have been estimated exploiting hadronic probes, using an extrapolation of data from antiprotonic atom x-rays~\cite{PhysRevLett.87.082501}.
From these data, the value of the so-called neutron skin, $\Delta R_{np}\equiv R_{n} - R_{p}$, has been
measured for a number of elements, from which the extrapolated neutron skin value for each nucleus was found assuming a linear dependence on the asymmetry parameter, $I= (N-Z)/A$, where $A$ is the mass number, leading to the empirically fitted function 
\begin{equation}
\Delta R_{np}^{\mathrm{had}}(\mathcal{N}) =(-0.04\pm 0.03)+\left(1.01\pm 0.15\right)\, I \textrm{ fm} .
\label{eq:eqhadronic}
\end{equation}
Using the latter equation, the extrapolated value of the neutron skin of $^{133}\text{Cs}$ is $\Delta R_{np}^{\mathrm{had}}(\mathrm{^{133}Cs})=0.13(4)\textrm{ fm}$, that combined with the very well known value of $R_{p}(^{133}\text{Cs})=4.807(1) \textrm{ fm}$ at the time~\cite{JOHNSON1985405}, gave a value of $R_n(^{133}\text{Cs})=4.94(4) \textrm{ fm}$ and a correction to $Q_W^{\mathrm{^{133}Cs},\, \mathrm{PDG}}$ explicitly visible in Table IV of Ref.~\cite{Dzuba:2012kx}.  
However, these determinations of the neutron skin with hadronic measurements are known to be affected, unlike electroweak measurements, by considerable model dependencies and uncontrolled approximations~\cite{Thiel:2019tkm}. In this paper, we revisit the determination of $Q_W^{\mathrm{^{133}Cs}}$ determining $R_n(^{133}\text{Cs})$ from a practically model-independent extrapolation from the recent average neutron rms radius of $^{208}\text{Pb}$ performed by the PREX-1 and PREX-2 experiments~\cite{Horowitz:2012tj,Adhikari:2021phr, Abrahamyan_2012,PREXII}, which exploit parity violating electron scattering on lead. 
Indeed, the PREX collaboration released a unique determination of the point neutron skin, the difference between the point\footnote{The physical proton and neutron radii
$R_{p,n}$ can be retrieved
from the corresponding point-radii
$R_{p,n}^{\text{point}}$
adding in quadrature the contribution of the
rms nucleon radius
$\langle r_{N}^2 \rangle^{1/2} \simeq 0.84 \, \text{fm}$,
that is considered to be approximately equal for the proton and the neutron. Namely, 
$R_{p,n}^2= (R_{p,n}^{\text{point}})^2 + \langle r_{N}^2 \rangle$.} 
neutron and proton rms radii $R_{p,n}^{\text{point}}$, that is equal to~\cite{Horowitz:2012tj,Adhikari:2021phr, Abrahamyan_2012,PREXII} 
\begin{equation}
    \Delta R_{np}^{\mathrm{point}} (^{208}\text{Pb}) \equiv R_{n}^{\mathrm{point}} - R_{p}^{\rm{point}} = 0.283 (71)\,\mathrm{fm}.
\end{equation}
We note that, this value is significantly larger with respect to the one that could be retrieved using the extrapolation in Eq.~(\ref{eq:eqhadronic}), corresponding to $\Delta R_{np}^{\mathrm{had}}(^{208}\text{Pb}) = 0.17(4)\,\mathrm{fm}$. Given that the PREX measurement is basically model independent and thus more reliable, this large discrepancy motivated us to discard the determination of $R_n(^{133}\text{Cs})$ from hadronic probes in favor of the usage of electroweak probes. 
\begin{figure}[!b]
\centering
\includegraphics*[width=\linewidth]{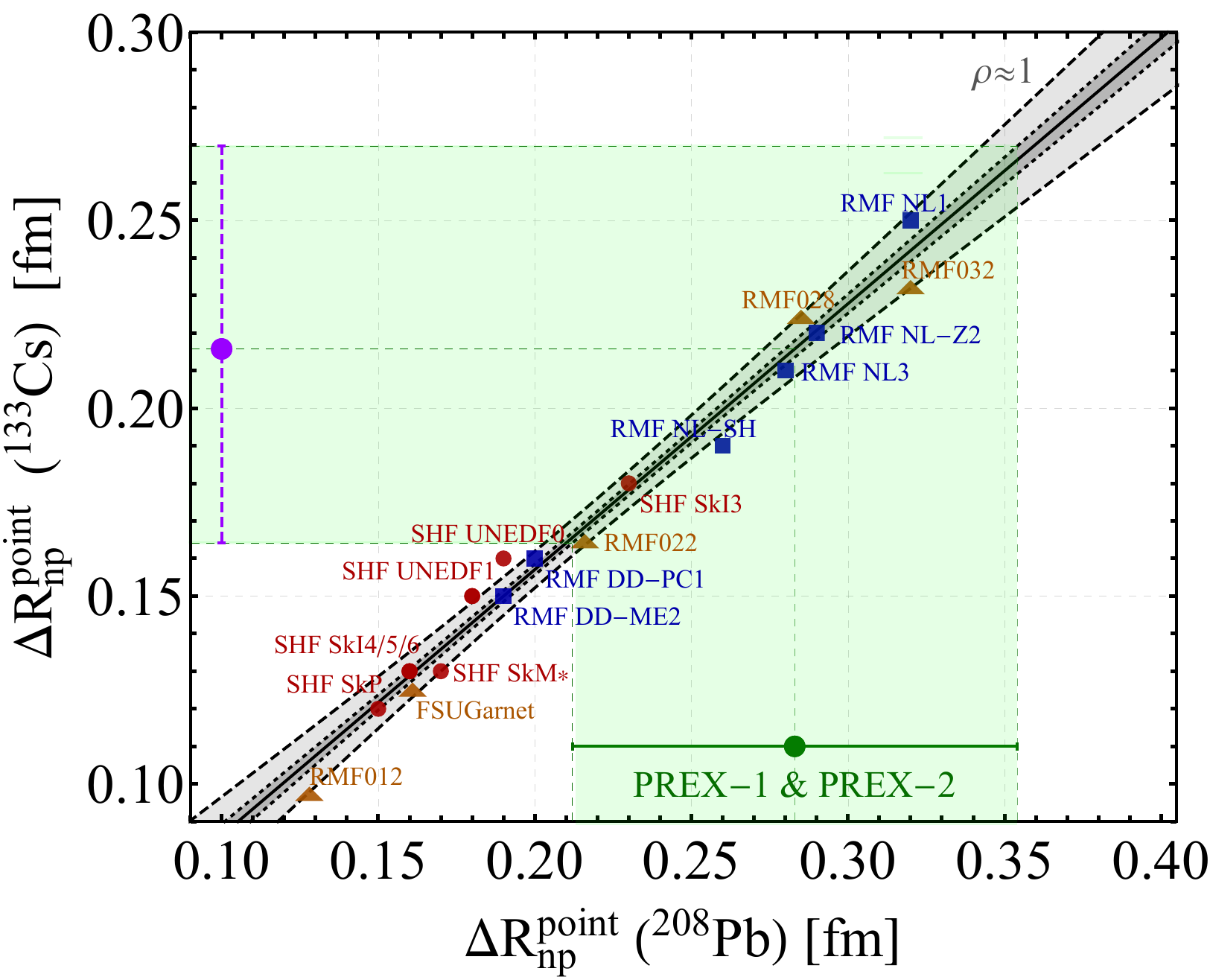}
\caption{ \label{fig:neutron_skin}
Point neutron skin predictions for $^{208}\text{Pb}$ and $^{133}\text{Cs}$ according to different models (red circles~\cite{Dobaczewski:1983zc,Bartel:1982ed,Kortelainen:2011ft,Kortelainen:2010hv,Chabanat:1997un,Reinhard:1995zz}, orange triangles~\cite{Hernandez:2019hsk,Yang:2019pbx,Chen:2014sca,Chen:2014mza}, and blue squares~\cite{Sharma:1993it,Bender:1999yt,Lalazissis:1996rd,Reinhard:1986qq,Niksic:2008vp,Niksic:2002yp}). 
A linear fit is superimposed in solid black, where the $1\sigma$ and $3\sigma$ constraints are also shown by the dark-gray dotted and light-gray dashed regions, respectively.
Constraints set by the combination of PREX-1 and PREX-2~\cite{Horowitz:2012tj,Adhikari:2021phr, Abrahamyan_2012,PREXII} and the constraint on $\Delta R_{\mathrm{np}}^{\mathrm{point}} (^{133}\text{Cs})$ derived in this work are also shown by the green and purple point, respectively.}
\end{figure}%

To this purpose, in Fig.~\ref{fig:neutron_skin} we show the values of the point neutron skins of $^{133}\text{Cs}$,
and
$^{208}\text{Pb}$
obtained with various
nonrelativistic Skyrme-Hartree-Fock (SHF)~\cite{Dobaczewski:1983zc,Bartel:1982ed,Kortelainen:2011ft,Kortelainen:2010hv,Chabanat:1997un,Reinhard:1995zz} 
and
relativistic mean-field (RMF)~\cite{Sharma:1993it,Bender:1999yt,Lalazissis:1996rd,Reinhard:1986qq,Niksic:2008vp,Niksic:2002yp,Hernandez:2019hsk,Yang:2019pbx,Chen:2014sca,Chen:2014mza}
nuclear models.
A clear model-independent linear correlation~\cite{Yang:2019pbx, Zheng:2014nga, Sil:2005tg, PhysRevC.85.041302,Yue:2021yfx,cadeddu2021new} is present between the neutron skin of $^{208}\text{Pb}$ and $^{133}\text{Cs}$ within the nonrelativistic and relativistic models with different interactions, with a Pearson's correlation coefficient $\rho \simeq 0.999$, an angular coefficient equal to $0.707 \pm 0.023$ and intercept $0.016 \pm 0.005$~fm. Here we want to exploit this powerful linear correlation to translate the PREX-1 \& PREX-2 combined measurement of $\Delta R_{\mathrm{np}}^{\mathrm{point}} (^{208}\text{Pb})$ into a determination of $\Delta R_{\mathrm{np}}^{\mathrm{point}} (^{133}\text{Cs})$. We obtain
\begin{equation}
    \Delta R_{\mathrm{np}}^{\mathrm{point}} (^{133}\text{Cs}) = 0.22(5) \,\mathrm{fm}.
\end{equation}
Comparing it with the extrapolated value derived using hadronic probes, we note that the uncertainty is basically the same while the central value is almost doubled. 
This measurement can be in turn translated into a rather-precise and model-independent value of the physical neutron rms radius, exploiting the well-known value of the proton rms radius determined experimentally from muonic atom spectroscopy~\cite{Fricke:1995zz,Angeli:2013epw} and corrected following the procedure introduced in Refs.~\cite{Cadeddu:2018izq,Cadeddu:2020lky}, corresponding to
$R_{p}(^{133}\text{Cs})
=
4.821(5) \, \text{fm}$. We thus obtain 
\begin{equation}
   R_{n}(^{133}\text{Cs})
= 5.03(5) \, \text{fm}.
\label{RnCs}
\end{equation}
This value is also compatible with the phenomenological nuclear shell model estimation in Ref.~\cite{Hoferichter_2020} and can be used as an input for $Q_W^{\mathrm{^{133}Cs},\,\mathrm{exp}}$.\\
Experimentally, the weak charge of Cs is extracted from the ratio of the parity violating amplitude, $E_{\text{PNC}}$, to the Stark vector transition polarizability, $\beta$, and by calculating theoretically $E_{\rm PNC}$ in terms of $Q_W^{^{133}\text{Cs},\,\mathrm{SM}}$, leading to
\begin{align}
\label{QWeqformula}
Q_W^{^{133}\text{Cs},\,\mathrm{exp}} &= N_{\mathrm{Cs}} \left( \dfrac{{\rm Im}\, E_{\rm PNC}}{\beta} \right)_{\rm exp.}\cdot \nonumber \\
&\cdot \left( \dfrac{Q_W^{^{133}\text{Cs},\,\mathrm{SM}}}{N_{\mathrm{Cs}}\, {\rm Im}\, E_{\rm PNC} (R_n)} \right)_{\rm th.} \beta_{\rm exp.+th.}\,,
\end{align}
where $\beta_{\rm exp.+th.}$ and $(\mathrm{Im}\, E_{\rm PNC})_{\rm th.}$ are determined from atomic theory, and Im stands for imaginary part (see Ref.~\cite{Zyla:2020zbs}).
In particular, we use 
$({\rm Im}\, E_{\rm PNC}/{\beta})_{\rm exp} = (-3.0967 \pm 0.0107) \times 10^{-13} |e|/a_B^2$~\cite{Zyla:2020zbs}, where $a_B$ is the Bohr radius and $\beta_{\rm exp.+th.} = (27.064 \pm 0.033)\, a_B^3$~\cite{Zyla:2020zbs}. The imaginary part of $E_{\rm PNC}$ is where the dependence on the value of $R_n(^{133}\text{Cs})$ is encapsulated. Thus, we use
$({\rm Im}\, E_{\rm PNC})_{\rm th.}^{\rm w.n.s.}=(0.8995\pm0.0040)\times10^{-11}|e|a_B Q_W^{^{133}\text{Cs},\,\mathrm{SM}}/N_{\mathrm{Cs}}$~\cite{Dzuba:2012kx}, where we subtracted the correction called ``neutron skin," introduced to take into account the difference between $R_n$ and $R_p$ that is not considered in the nominal atomic theory derivation. Indeed, besides the usage of the value of $\Delta R_{np}^{\mathrm{had}}(^{133}\text{Cs})$, that we have shown to be quite model-dependent, another problem connected with this correction is that it was determined using the approximated formula in Eq.~(4.8) of Ref.~\cite{PhysRevA.65.012106}, that underestimates the correction for larger values of $\Delta R_{\mathrm{np}}$.
Here we remove this correction in order to be able to implement a new one with the value of $R_{n}(^{133}\text{Cs})$ just derived and avoiding the usage of an approximated formula.
The neutron skin corrected value of the weak charge is thus retrieved using the correcting term~\cite{Viatkina,Cadeddu:2019eta,Cadeddu:2018izq}
\begin{align}
\frac{\delta E^\mathrm{n.s.}_\mathrm{PNC}(R_{n}(^{133}\text{Cs}))}{E_\mathrm{PNC}^\mathrm{w.n.s.}}
=
\frac{N_\mathrm{Cs}}{Q_W^{^{133}\text{Cs},\,\mathrm{SM}}}\left(1-\frac{q_n(R_{n}(^{133}\text{Cs}))}{q_p}\right)
%\nonumber\\
%= -0.0036
,
\end{align}
where the factors $q_p$ and $q_n$ incorporate the radial dependence of the electron axial transition matrix element by considering the proton and the neutron densities in the
nucleus as functions of the radius $r$, $\rho_{p,n}(r)$. Namely,
\begin{equation}
    q_{p,n}=4\pi \int_0^\infty \rho_{p,n}(r) f(r) r^2\mathrm{d}r,
    \label{eq:qpn}
\end{equation}
where $f(r)$
is the matrix element of the electron axial current between the atomic
$s_{1/2}$ and $p_{1/2}$ wave functions inside the nucleus normalized to $f(0)=1$. The details of the calculation can be found in the Supplemental Material of Ref.~\cite{cadeddu2021new}.
The new experimental value of the weak charge of $^{133}\text{Cs}$ becomes
\begin{equation}
    Q_W^{^{133}\text{Cs},\,\mathrm{exp}} = -72.94(43).
    \label{eq:APVour}
\end{equation} 
This result can be compared to the current one presented in Eq.~(\ref{APVPDG}). The uncertainty is practically the same and the central value is only marginally shifted. However, the main advantage is that it is derived from a determination of $R_{n}(^{133}\text{Cs})$ that is coming solely from electroweak probes with less assumptions.
\\

The measurements of $Q_W$ in Eqs.~(\ref{eq:APVour}) and~(\ref{Qprotonexp}) can be used to set limits on the available phase space for the $Z_d$ model. Indeed, the presence of a $Z_d$ mediator would change the experimental values of $Q_W$.
More precisely, adopting the substitutions described before, the proton weak charge expression becomes
\begin{align}
    Q_W^{p,Z_d}=-2\rho_d\  g^{ep}_{AV}(\kappa_d\sin^2\theta_W)\ \Big(1-\frac{\alpha}{2\pi}\Big),
\end{align}
where, in the case of polarized electron scattering experiments, such as for the measurement of the proton weak charge, the propagator term inside Eqs.~(\ref{rhod}) and (\ref{kappad}) becomes ~\cite{BOUCHIAT198373,Bouchiat_2005}
\begin{equation}
   f\Big(\frac{Q^2}{\mzd^2}\Big)=\frac{\mzd^2}{\mzd^2+Q^2},
\end{equation}
where $Q^2$ is the typical momentum transfer of the experiment.\\
Similarly, the expression for the cesium weak charge is
\begin{align}
    Q_W^{\mathrm{^{133}Cs},\ Z_d}
    =
    &-2\rho_d \Big[Z_{\mathrm{Cs}}(g^{ep}_{AV}(\kappa_d\sin^2\theta_W)+0.00005)\nonumber
    &&\\
    &+ N_{\mathrm{Cs}}(g^{en}_{AV}+0.00006)\Big]\Big(1-\frac{\alpha}{2\pi}\Big).&&
\end{align}
In the case of parity violation in heavy atoms, such as for cesium, the propagator assumes a different form due to the nuclear structure. In particular, for $^{133}\mathrm{Cs}$ it becomes $f(Q^2/\mzd^2)=K(^{133}\mathrm{Cs})$, as described in Refs.~\cite{BOUCHIAT198373,Bouchiat_2005}. For example, $K(^{133}\mathrm{Cs}) \simeq 0.5$ for masses of the $Z_d$ boson of the order of the typical momentum transfer of APV, $Q\approx 2.4\ \mathrm{MeV}$, while $K(^{133}\mathrm{Cs}) \simeq 0.83,1$ for $m_{Z_d} \simeq 20,100\ \mathrm{MeV}$.\\

\begin{figure}[!t]
\centering
\includegraphics*[width=\linewidth]{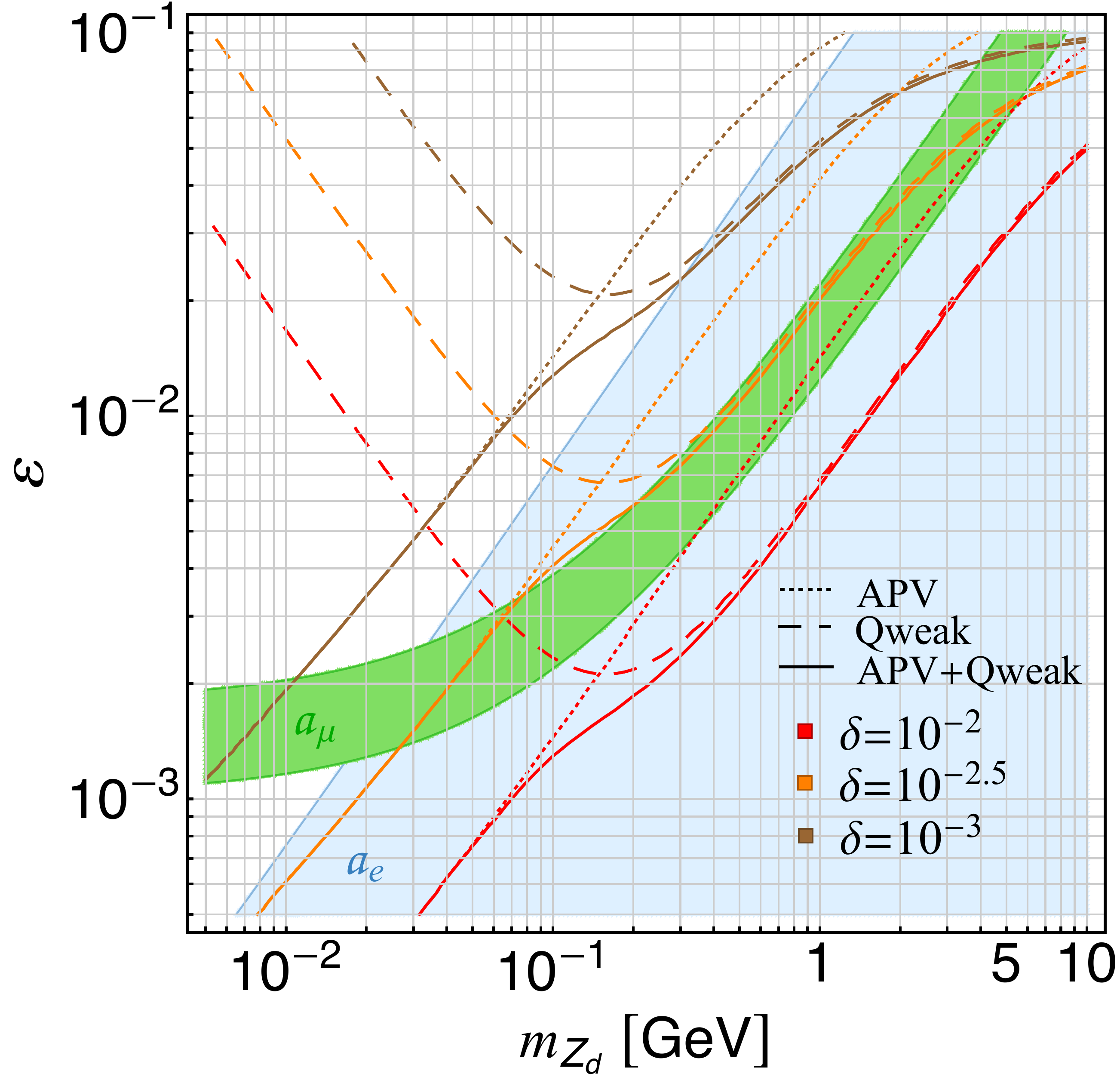}
\caption{ \label{fig:chi2_marciano} Limits at 90\% CL in  the  plane  of $m_{Z_d}$ and $\varepsilon$, for both ${\rm Q_{weak}}$ (dashed line) and APV (dotted line) experiments, and also their combination (solid line), for different values of $\delta$ as depicted in the label. The green band and the light blue area are the favored regions at 90\% CL needed to explain the anomalous magnetic moment of the muon and of the electron, respectively.
}
\end{figure}

In order to determine information on $\varepsilon$, $\delta$ and $m_{Z_d}$, we performed several fits with the common least-squares function
\begin{align}
\chi^2_i
=
\dfrac{
(X^{\rm exp}_i
-
X^{\mathrm{th}}_i(\varepsilon,\, \delta,\,m_{Z_d}))^2
}{\sigma_{i}^2}
\,,
\label{chi2}
\end{align} 
where $i$ stands for ${\rm Q_{weak}},\, {\rm APV},\, {a_\mu}$, and ${a_e}$, such that $X^{\rm exp}=\{ Q_W^{p,\,\mathrm{exp}},\, Q_W^{^{133}\text{Cs},\, \mathrm{exp}},\, a_\mu^{\rm exp},\, a_e^{\rm exp}\}$,  $X^{\rm th}=\{Q_W^{p,\,Z_d}, Q_W^{^{133}\text{Cs},\,Z_d},\,a_\mu^{Z_d},\, a_e^{Z_d}\}$ and $\sigma_{i}$ are the corresponding experimental and theoretical uncertainties summed in quadrature. 
In Fig.~\ref{fig:chi2_marciano} we show the limits or allowed regions at 90\% confidence level (CL) in  the  plane  of $m_{Z_d}$ and $\varepsilon$ for different values of $\delta$. In particular, we show the limits of APV, ${\rm Q_{weak}}$ and their combination. Moreover, we also show the 90\%~CL favored regions for the explanation of the muon and electron anomalous magnetic moments. 
We note that the ability to exclude the $a_\mu$ and $a_e$ interpretations under the $Z_d$ model depends strongly on the value of $\delta$ chosen. Namely, for $\delta>10^{-2}$ the entire $\Delta a_\mu$ discrepancy is completely ruled out, not only by the combined result but also by the APV only limit. 
Other experiments that are also sensitive to $Z_d$ bosons are those able to measure rare flavor-changing weak neutral-current decays of $K$ and $B$ mesons, like $K^{\pm} \to \pi^{\pm} Z_d$, induced by quark transition amplitudes such as $s\to d Z_d$ and $b\to s Z_d$~\cite{Davoudiasl:2014kua, Essig:2013vha,Izaguirre:2013uxa,Pospelov:2007mp}.  Similarly, Higgs boson decays to $Z Z_d$ bosons~\cite{Davoudiasl:2012ag,Davoudiasl_2015}, induced by $Z-Z_d$ mass mixing, are also sensitive to it.
In both cases, the constraints obtained depend on the assumed branching fraction (BF) of the $Z_d$ boson decay and on its mass. Indeed, if only SM particles are lighter than the $Z_d$ boson, the latter could decay into pairs of charged leptons or hadrons, leaving a visible signature in detectors\footnote{Although this depends on the $Z_d$ lifetime, otherwise the $Z_d$ boson may escape and decay outside the detector acceptance making it impossible to reconstruct its decay products.}, or into a pair of neutrinos, resulting in missing energy. Instead, if it exists at least one dark-matter particle whose mass is such that $2 m_\chi < m_{Z_d}$, the $Z_d$ boson decays preferentially into dark-matter, depending on the assumed coupling $\alpha_D$. 
In the mass range $30\lesssim m_{Z_d} \lesssim 300$ MeV the dominant constraints arise from rare kaon decays.
Indeed, the BNL E949 experiment combined with the E787 results put severe constraints on $K^+ \to \pi^+ +$invisible~\cite{Artamonov:2009sz}, that however can be significantly relaxed in case of $\delta \neq 0$ thanks to a cancellation that may occur between kinetic and $Z - Z_d$ mass mixing~\cite{Davoudiasl:2014kua}.
It is worth to mention that the two favored regions determined for the magnetic moments in  Fig.~\ref{fig:chi2_marciano} do not depend significantly on the value of $\delta$, at least for the small values tested in this work, since the dominant contribution is the one induced by the term related to the kinetic mixing parameter $\eps$ in Eq.~(\ref{eq:5}).\\

\begin{figure}[t]
\centering
\includegraphics*[width=0.97\linewidth]{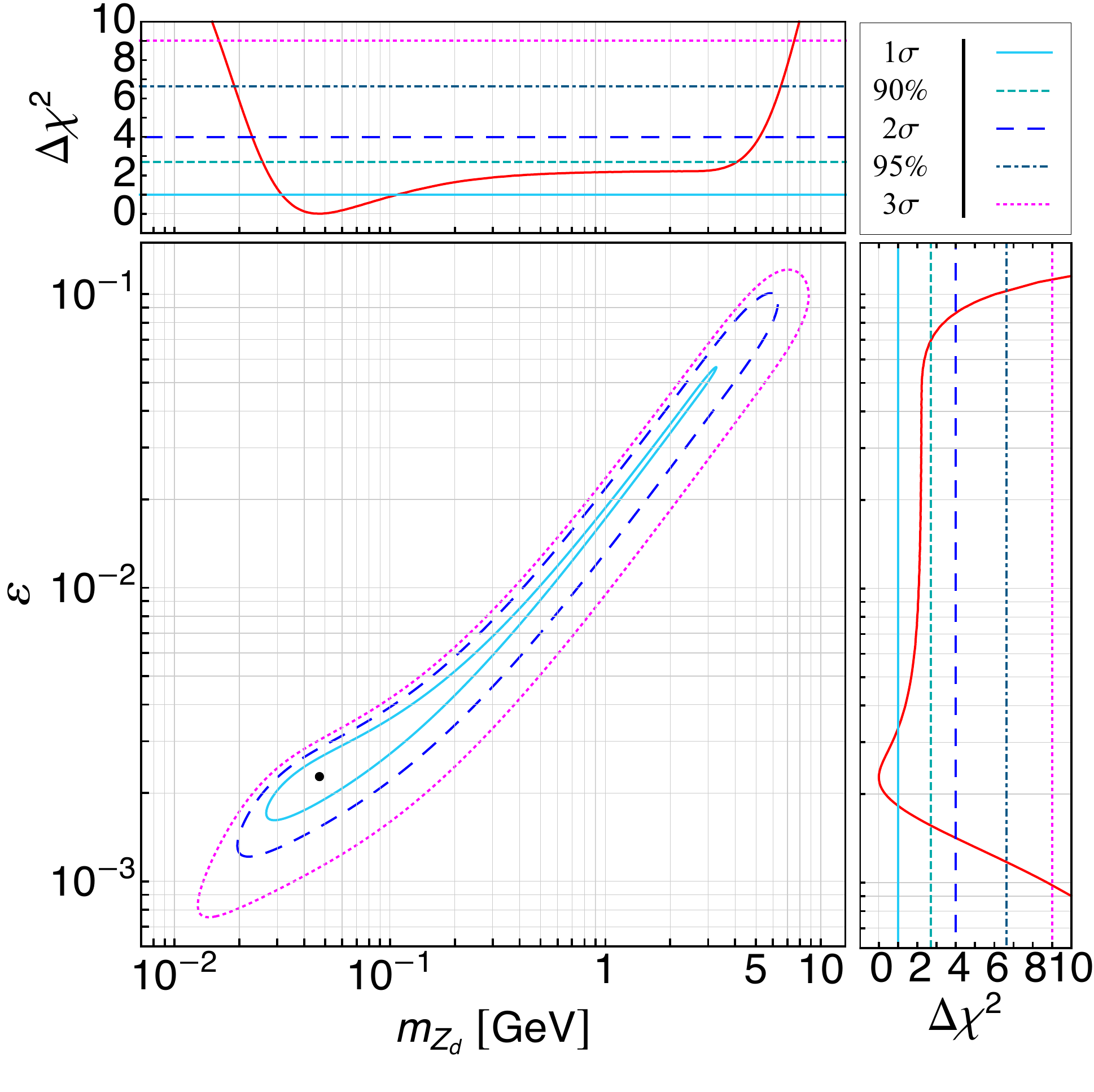}
\caption{ \label{fig:Combined_fit} 
Contours  at  different  CL  of  the  allowed  regions in  the  plane  of $m_{Z_d}$ and $\varepsilon$,  together  with  their marginalizations, obtained from the combined fit of the ${\rm Q_{weak}}$, APV, $a_\mu$ and $a_e$ experimental results. The best fit result is indicated by the black dot.
}
\end{figure}

In Fig.~\ref{fig:chi2_marciano} it is possible to see that, for given values of $\delta$, $m_{Z_d}$ and $\varepsilon$, there is an overlap between all the different experimental constraints. To better highlight it, we performed a combined fit by summing all the four $\chi^2$'s in Eq.~(\ref{chi2}). In order to remove the ambiguity on $\delta$, we marginalized the result over this parameter. 
In Fig.~\ref{fig:Combined_fit} we show the $1\sigma$, $2\sigma$, and $3\sigma$ CL contours in the plane of $m_{Z_d}$ and $\varepsilon$, as well as the best fit result corresponding to a minimum $\chi_{\mathrm{min}}^2=0.007$.
\begin{figure}[ht]
\centering
\includegraphics*[width=\linewidth]{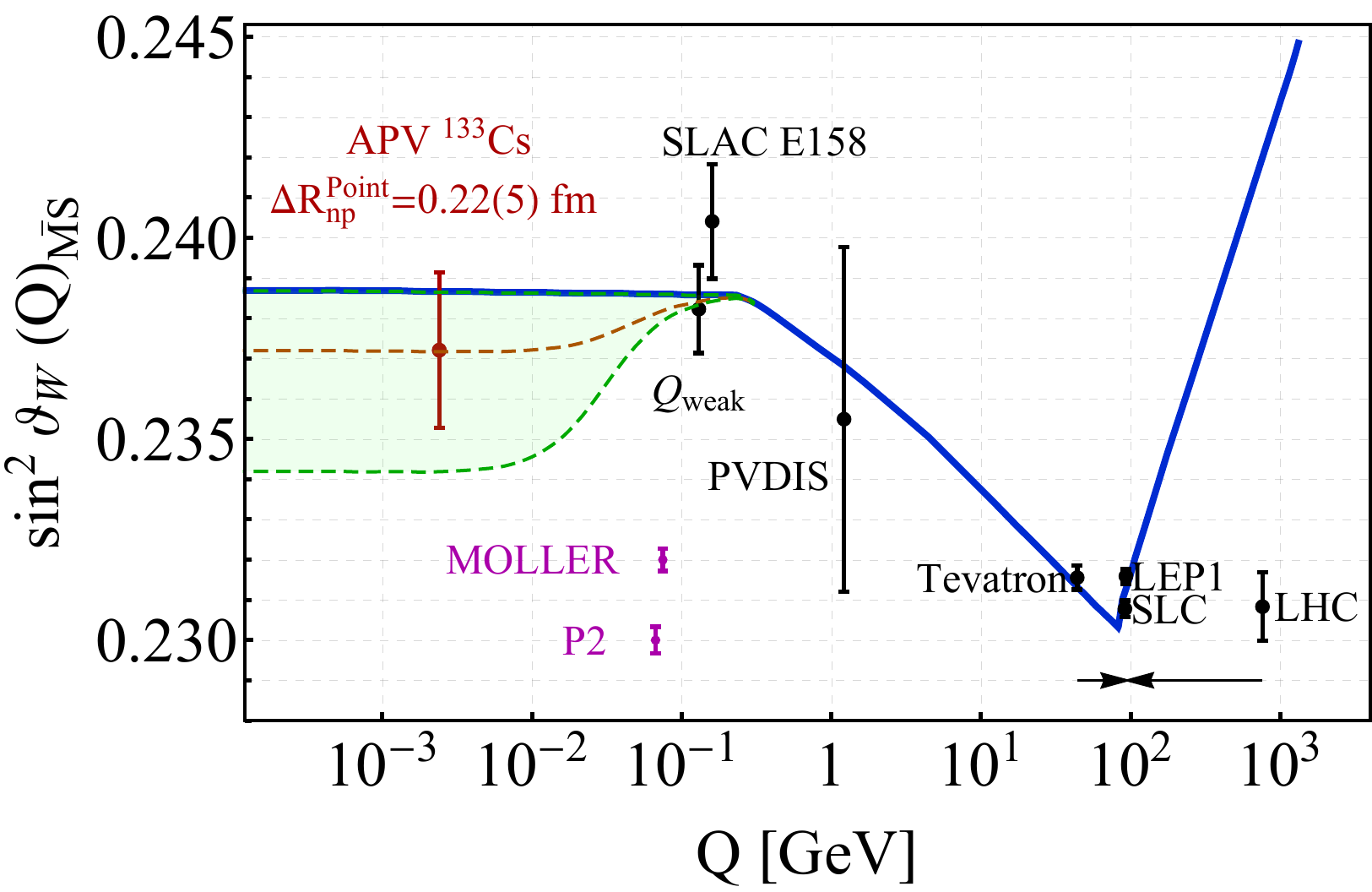}
\caption{ \label{fig:running}
Variation of $\sin^2 \vartheta_{\text{W}}$ with energy scale $Q$. The SM prediction is shown as the solid blue curve, together with
experimental determinations in black~\cite{Tanabashi:2018oca,Wood:1997zq,Dzuba:2012kx,Anthony:2005pm,Anthony:2005pm,Wang:2014bba,Zeller:2001hh, Androic:2018kni} and future projections in violet~\cite{Becker:2018ggl,Benesch:2014bas} with a central value shown at an arbitrary position. The result derived in this paper for APV on cesium is shown in red.
With the dashed red and green lines we indicate the best fit result and the $\pm 1\sigma$ variations, respectively, for the running of $\sin^2 \vartheta_{\text{W}}$ in the presence of a $Z_d$ boson as described in the paper.  }
\end{figure}
For completeness, when marginalizing in turn over the other two parameters, we get the following results for $m_{Z_d}$, $\varepsilon$ and $\delta$ at $1\sigma$ CL
\begin{align}
    m_{Z_d} &= 47{}^{+61}_{-16} \, \mathrm{MeV},\\
    \varepsilon &= 2.3{}^{+1.1}_{-0.4} \times 10^{-3},\\
    \delta &< 2 \times 10^{-3}.
\end{align}
Using these best fit values\footnote{The best fit value of $\delta$ is $7.9\times 10^{-4}$, see Ref.~\cite{refsupp} for additional information.} and their $1\sigma$ ranges, in Fig.~\ref{fig:running} we show how the running of $\sin^2 \vartheta_{\text{W}}$ changes at low energies due to the contribution of a $Z_d$ boson. Clearly, further measurements of $\sin^2 \vartheta_{\text{W}}$ in the low energy sector, as those coming from the P2~\cite{Becker:2018ggl,Dev:2021otb} and MOLLER~\cite{Benesch:2014bas} experiments, from the near DUNE detector~\cite{deGouvea:2019wav}, the exploitation of coherent elastic neutrino scattering in atoms~\cite{Cadeddu_2019} and nuclei~\cite{Cadeddu:2020lky,Fernandez_Moroni_2021,Ca_as_2018} and finally from future atomic parity violation with francium, radium and rubidium~\cite{RevModPhys.90.025008, Roberts_2015}  would be really powerful for further constraining such a model.
\begin{figure}[!t]
\centering
\includegraphics*[width=\linewidth]{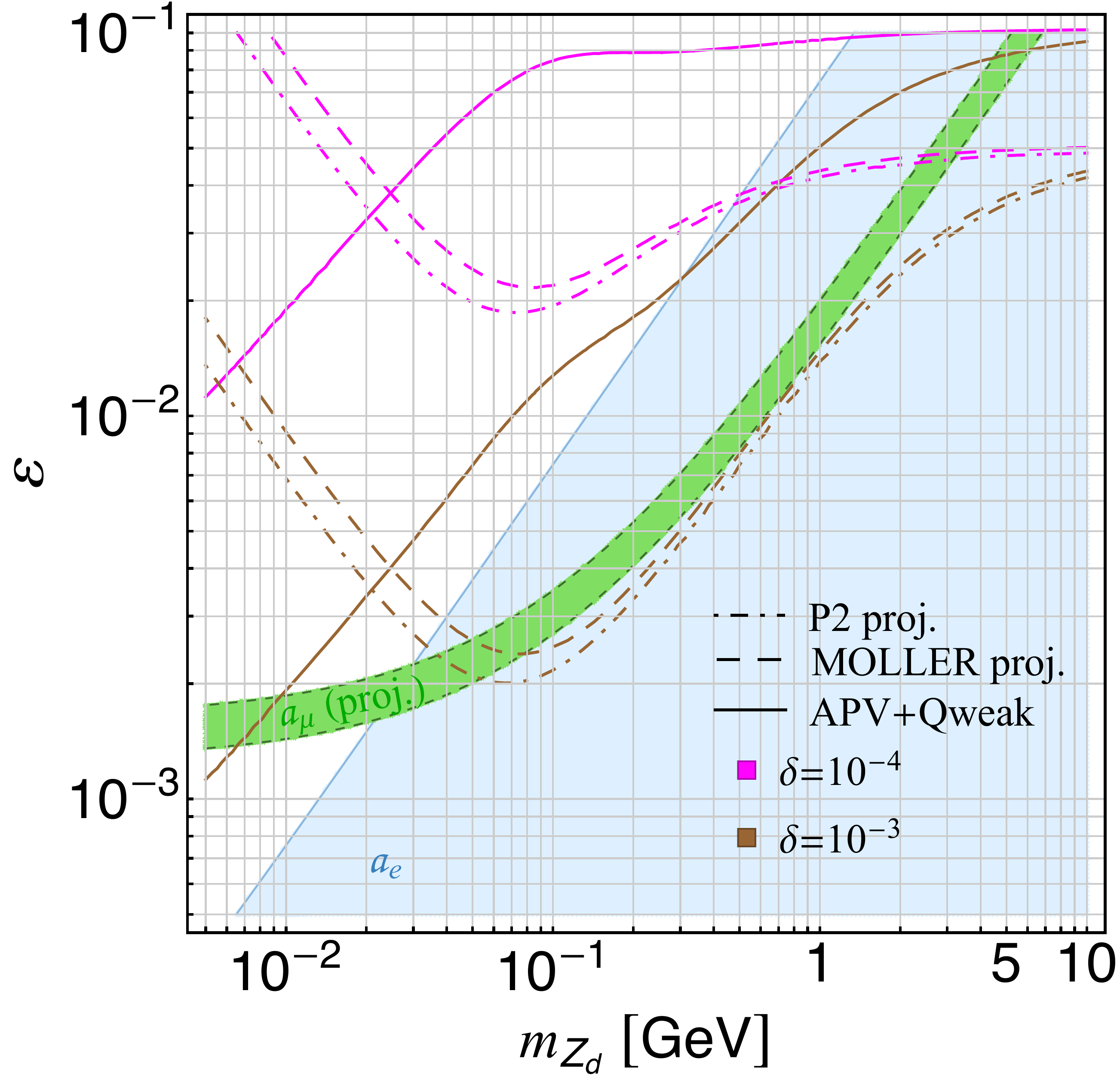}
\caption{ \label{fig:chi2_marciano_future} Limits at 90\% CL in  the  plane  of $m_{Z_d}$ and $\varepsilon$, for the combination of the current ${\rm Q_{weak}}$ and APV experiments (solid), the projected MOLLER (dashed) and P2 (dot-dashed) proposed experiments, for different values of $\delta$ as depicted in the label.
The green band and the light blue area are the favored regions at 90\% CL needed to explain the projected and current anomalous magnetic moment of the muon and of the electron, respectively.
}
\end{figure}
To highlight the near future prospects that can be achieved thanks to upcoming results from MOLLER and P2, considering the SM value for the central value, as well as an improved measurement of $a_\mu$ with half of the uncertainty in~Eq.~(\ref{amuexp}), we show in Fig.~\ref{fig:chi2_marciano_future} the limits at 90\% CL in  the  plane  of $m_{Z_d}$ and $\varepsilon$ for different values of $\delta$. As clearly visible, P2 and MOLLER will allow to exclude a large portion of the $a_\mu$ band already for values of $\delta$ as small as $10^{-3}$. See Ref.~\cite{refsupp} for additional information.
\\

In summary, in this paper we studied a possible $U(1)_d$ extension of the SM that implies the presence of a sub-GeV-scale vector $Z_d$ mediator. The existence of this additional force would modify the experimental values of the muon and electron anomalous magnetic moments as well as the measurements of the proton and cesium weak charge, performed so far at low-energy transfer. Motivated by the recent determination of the muon anomalous magnetic moment performed at Fermilab, we derived the constraints on such a model obtained from the aforementioned experimental measurements and by their combination. Before to do so, we revisited the determination of the cesium $Q_W$ from the atomic parity violation experiment, which depends critically on the value of the average neutron rms radius of $^{133}\text{Cs}$, by determining the latter from a practically model-independent extrapolation from the recent average neutron rms radius of $^{208}\text{Pb}$ performed by the PREX-2 Collaboration. From a combined $\chi^2$ fit we obtain rather precise limits on the mass and the kinetic mixing parameter of the $Z_d$ boson, namely $m_{Z_d} = 47{}^{+61}_{-16} \, \mathrm{MeV}$
and $\varepsilon = 2.3{}^{+1.1}_{-0.4} \times 10^{-3}$, when marginalizing over the $Z-Z_d$ mass mixing parameter $\delta$.

%\begin{acknowledgements}
%\end{acknowledgements}

%\nocite{*}
\bibliographystyle{apsrev4-1}
\bibliography{bibAPV}

\clearpage

\end{document}

\appendix

\begin{center}
 \bf \large Supplemental Material
\end{center}
\vspace*{0.2cm}

In the following appendix we provide additional details on the near future prospects for constraining a $Z_d$ model using further foreseen measurements of $\sin^2 \vartheta_{\text{W}}$ and an improved determination of $a_\mu$.

\begin{figure}[b]
\centering
\includegraphics*[width=0.49\linewidth]{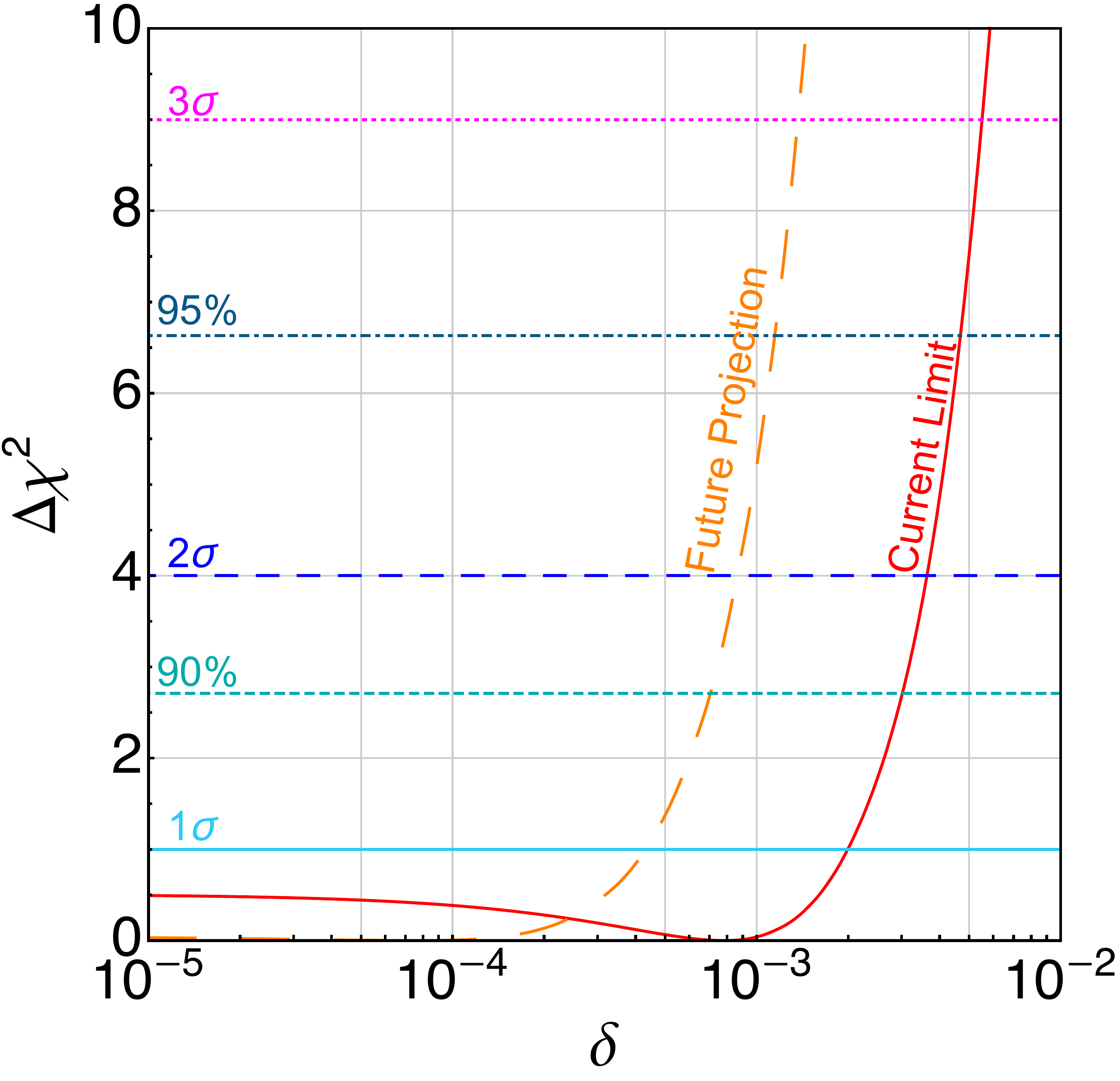}
\includegraphics*[width=0.47\linewidth]{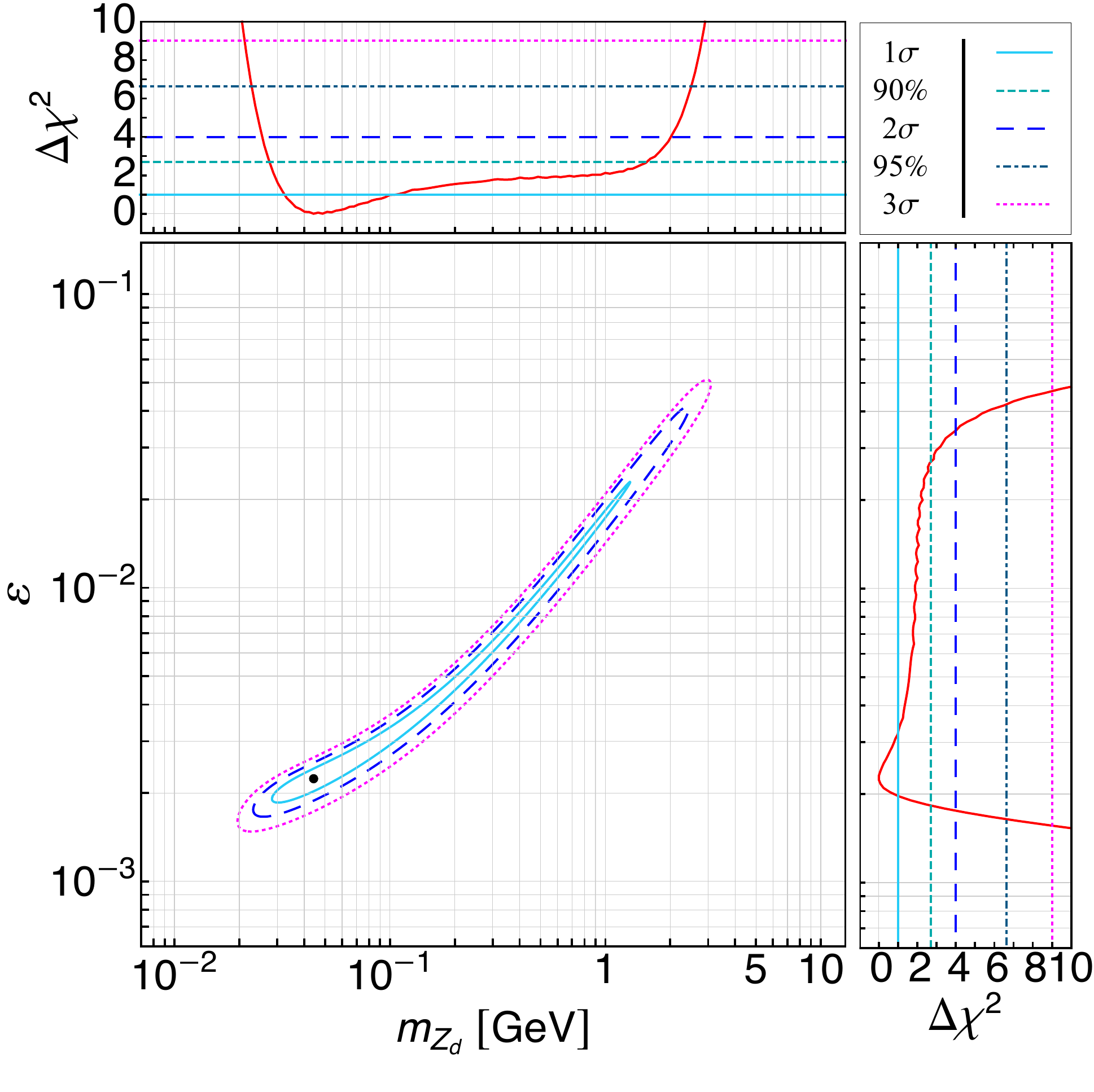}
\caption{ \label{fig:future_proj} 
(Left) Marginalizations at different CL for the $\delta$ parameter obtained from the combined fit of  ${\rm Q_{weak}}$, APV, $a_e$ and $a_\mu$ experimental results (in solid red) and including the projected sensitivity of P2 and MOLLER experiments as well as for $a_\mu$ (in dashed orange). See the text for details.  (Right) Contours  at  different  CL  of  the  allowed  regions in  the  plane  of $m_{Z_d}$ and $\varepsilon$,  together  with  their marginalizations, obtained from the combined fit of  ${\rm Q_{weak}}$, APV, $a_e$ experimental results, the projections for P2 and MOLLER and future $a_\mu$ expected sensitivity. The best fit result is indicated by the black dot.
}
\end{figure}

Namely, we use the precision declared for the upcoming P2~\cite{Becker:2018ggl} and MOLLER~\cite{Benesch:2014bas} experiments, considering for the central value the SM predicted value at low momentum transfer and an experimental uncertainty of 1.5\% and 2.4\% on the $Q_W$ measurement, respectively. 
Moreover, we consider an improved measurement of $a_\mu$ with half of the uncertainty obtained in Ref.~\cite{PhysRevLett.126.141801}, as achievable by the Muon g-2 Collaboration with the inclusion of the already available data of Run2 and Run3.
In Fig.~\ref{fig:future_proj}~(left) we show 
the marginalizations at different CL for the $\delta$ parameter obtained from a combined fit with the inclusion of the aforementioned projections and the current ${\rm Q_{weak}}$, APV, and $a_e$ experimental results. Clearly, it will be possible to improve the current limits on $\delta$ by almost one order of magnitude. \\
In Fig.~\ref{fig:future_proj}~(right) we show the contours  at  different  CL  of  the  allowed  regions in  the  plane  of $m_{Z_d}$ and $\varepsilon$,  together  with  their marginalizations, obtained including the future projections of P2, MOLLER and $a_\mu$ and marginalizing over $\delta$, with a minimum $\chi^2$ of $(\chi^2_{\mathrm{min}})^{\mathrm{proj}} \simeq 0.5$. A clear reduction of the $3\sigma$ contour is visible.
For completeness, when marginalizing in turn over the other two parameters, we get the following results for $m_{Z_d}$, $\varepsilon$ and $\delta$ at $1\sigma$ CL
\begin{align}
       m_{Z_d}^{\mathrm{proj}} &= 44{}^{+63}_{-12} \, \mathrm{MeV},\\
    \varepsilon^{\mathrm{proj}}&= 2.2{}^{+1.0}_{-0.3} \times 10^{-3},\\
    \delta^{\mathrm{proj}} &< 4 \times 10^{-4}.
\end{align}

\bibliographystyle{apsrev4-1}
\bibliography{supplement}